# Occupation-Driven emission asynchronous as a Fundamental Constraint on Solid-State Attosecond Pulses

Qing-Guo Fan[1,2], Kang Lai[1], Wen-hao Liu[1], Lin-Wang Wang[1,2,*], Jun-Wei Luo[1,2,**], Zhi Wang[1,†]

[1] State Key Laboratory of Semiconductor Physics and Chip Technologies, Institute of Semiconductors, Chinese Academy of Sciences, Beijing 100083, China

[2] Center of Material Science and Optoelectronics Engineering, University of Chinese Academy of Sciences, Beijing 100049, China

***Abstract***: *A newly analytic occupation-resolved theory capturing the temporal structure of attosecond pulses (APs) is derived. We validate it with real-time time-dependent density functional theory and show remarkable temporal confinement of APs with laser intensity in solid state. Using a simplified field-driven electron excitation together with a generalized pre-acceleration picture, the interband emission timing demonstrate intrinsically temporal mismatched with field-synchronous intraband radiation, leading to a nonmonotonic dependence of attosecond pulse width on laser intensity. Our findings not only shed light on the microscopic mechanisms behind solid-state high harmonic generation (HHG), but also establish the fundamental time-domain constraint on solid-state APs independent of material damage thresholds.*

**Corresponding authors:**

* lwwang@semi.ac.cn

** jwluo@semi.ac.cn

† wangzhi@semi.ac.cn

## I. INTRODUCTION

Solid-state high-harmonic generation (HHG) has emerged as a promising platform for producing attosecond pulses, combining sub-cycle temporal resolution with high efficiency and material tunability [1–5]. In close analogy to gas-phase HHG, it is widely assumed that the higher cutoff energy, the better temporal confinement of synthetic attosecond bursts [6,7]. Basing on the linear dependence of cutoff on field strength [8–12], searching for high damage threshold solid materials have therefore become a focal point of current researches [2,8,9,11–14]. Does this frequency-domain perspective is the only strategy to optimize solid-sate attosecond emission?

Another intriguing view is to utilize the underlying physical mechanism. Unlike gases, HHG in solids arises from the coexistence of two distinct microscopic mechanisms: intraband emission driven by carrier acceleration within dispersive bands, and interband emission associated with coherent electron-hole dynamics [9,10,15–17]. Recent experiments and simulations have reported attosecond bursts from intraband and interband emission, with durations ranging from several hundred attoseconds to the few-hundred-attosecond regime [4,15,18–20].

While the two channels are often discussed in terms of their spectral contributions, their relative timing in the time domain has received little attention. This omission is critical because attosecond pulses are intrinsically time-domain objects. Even when the harmonic spectrum is sufficiently broad, the temporal confinement of the emitted bursts depends on whether different emission channels radiate synchronously. If intraband and interband emission are not temporally aligned, their superposition can fundamentally limit attosecond pulse compression, irrespective of spectral bandwidth or material damage thresholds.

In this letter, we demonstrate an overlooked mechanism that solid-state attosecond emission is intrinsically constrained by carrier-occupation dynamics. By deriving an occupation-resolved theory of different emission channel, and interpreting it within intuitive physical picture, we show that interband radiation is generically delayed relative to the driving field, in sharp contrast to strictly field-synchronous intraband emission. Motivated by the promising potential of crystalline silicon for attosecond applications [3,21–24], we take bulk silicon as prototype, and find this intrinsic temporal mismatch leads to a nonmonotonic dependence of attosecond pulse width on laser intensity within the framework of real-time time-dependent functional theory. Our results establish carrier occupation as a fundamental time-domain limitation on attosecond pulse compression in solids.

## II. METHODS

**Calculating HHG, APs with rt-TDDFT.** To compute HHG spectra and APs within rt-TDDFT, we solve the time-dependent Kohn-Sham (TDKS) equations to obtain the Kohn-Sham orbital $\Psi_i(\boldsymbol{r},t)$, expanded in an adiabatic eigenstate basis [25]. Here, $i$ denotes the band index. The laser potential is considered in velocity gauge, and the TDKS equation reads:

$$\mathrm{i}\frac{\partial}{\partial t}\Psi_i(\boldsymbol{r},t) = \left(\frac{1}{2}(-i\nabla + \boldsymbol{A}^2) + v_{ext}(\boldsymbol{r},t) + v_{HF}[n(\boldsymbol{r},t)] + v_{xc}[n(\boldsymbol{r},t)]\right)\Psi_i(\boldsymbol{r},t) \quad (1)$$

Here, $v_{ext}$ is the electron-ion potential, $v_{HF}$ is the Hartree part of the Coulomb electron-electron interaction, $v_{xc}$ is the exchange-correlation potential, $\boldsymbol{A}$ is the laser vector potential, which is simulated in dipole approximation [26]: $\mathbf{A}(\mathbf{r},\mathrm{t}) \approx \mathbf{A}(\mathrm{t}) = A_0 sin^2(t\pi/\delta)\sin(\omega_0 \mathrm{t})$, where $A_0$ is the peak potential, $\delta$ is the total elapsed time, $\omega_0$ is the carrier angular frequency, the electric field $\mathbf{E}$ can be derived from the relationship: $\mathbf{E}(\mathrm{t}) = -\partial\boldsymbol{A}(t)/\partial t$.

We can get the time-dependent electron spatial density by $\mathrm{n}(\mathbf{r},\mathrm{t}) = \sum_i |\Psi_i(\boldsymbol{r},t)|^2$. The microscopic current $\mathbf{j}(\mathbf{r},\mathrm{t})$ can be computed as:

$$\boldsymbol{j}_i(\boldsymbol{r},t) = \frac{1}{2}(\langle\Psi_i(\boldsymbol{r},t)|-i\nabla + \boldsymbol{A}|\Psi_i(\boldsymbol{r},t)\rangle) + c.c. \quad (2)$$

where $c.c.$ denotes the complex conjugate. The macroscopic current is obtained by integrating the microscopic current over the system volume $\Omega$, $J(t) = \int_\Omega d^3\boldsymbol{r}\,\boldsymbol{j}_i(\boldsymbol{r},t)$.

The attosecond pulses (APs) in time domain and corresponding high harmonics generation (HHG) spectrum are obtained from squaring of the modulus and taking the

Fourier transform ($\mathcal{FT}$) of the first derivative of time-dependent current:

$$\mathrm{APs(t)} \propto \left|\frac{dJ(t)}{dt}\right|^2 \tag{3}$$

$$\mathrm{HHG}(\omega) \propto \left|\mathcal{FT}\left[\frac{dJ(t)}{dt}\right]\right|^2 \tag{4}$$

**Processing and characterizing signal.** The time-frequency analysis is performed by continuous wavelet transform [27]:

$$\mathrm{W}(\tau, \mathrm{s}) = \left|\frac{1}{\sqrt{s}} \int_{-\infty}^{\infty} dt \left[\frac{dJ(t)}{dt} \cdot \Phi^*\left(\frac{t-\tau}{s}\right)\right]\right|^2 \tag{5}$$

where $\tau, s$ is the translation coefficient and scaling coefficient, respectively. $\Phi^*(t)$ is the father wavelet function, the corresponding mother wavelet function is complex Morlet wavelet, which is given by: $\Phi(\mathrm{t}) = 1/\sqrt{\pi B} \cdot \exp\left(-t^2/B\right) \cdot \exp\left(j2\pi Ct\right)$, where $\mathrm{B}$ is the bandwidth (here $B = 1$), $C$ is the center frequency (chosen by proper parameters) and $\mathrm{j}$ is the imaginary unit.

The attosecond bursts ($\mathrm{AB}$) can be extracted from the coherent superstition of consecutive harmonics within a selected spectral window [7]:

$$\mathrm{AB(t)} = \left|\sum_{\omega=\omega_i}^{\omega_f} e^{i\omega t}\, \mathcal{FT}\left[\frac{dJ(t')}{dt'}\right]\right|^2 \tag{6}$$

where $\omega_i$ and $\omega_f$ determine the energy window to synthesize the $\mathrm{AB}$.

**Calculation of excited electron numbers.** The total number of excited electrons in band $n$ is obtained by projecting the time-dependent wavefunction $|\Psi_{n,k}(t)\rangle$ onto the ground-state Bloch wavefunctions $|\Psi_{n',k}^{GS}(t)\rangle$.

$$N_{ex}(t) = N_e(t) - \frac{1}{N_k} \sum_{n,n' \in occ} \sum_{k \in BZ} \left|\langle \Psi_{n,k} | \Psi_{n',k}^{GS} \rangle\right|^2 \tag{7}$$

where $N_e$ denotes the total number of electrons in the system, $N_k$ is the total number of sampling points in the Brillouin zone, and the indices $n$ and $n'$ run over all bands.

The addition simulation and derivation [2,10,25,28–41] are shown in Supplementary Materials.

## III. RESULTS and DISCUSSION

We start from the single-particle density matrix $\rho_k$[1] (see the Appendix B for details) and obtain, from the time-dependent Liouville equation,

$$\frac{d}{dt}\,\rho_k^{int}(t) = -\frac{i}{\hbar}\left[{H_{I,k}}^{int}, \rho_k^{int}(t)\right] \tag{1}$$

Where $H_{I,k}^{\mathrm{int}}$ and $\rho_k^{\mathrm{int}}(t)$ denote the laser–matter Hamiltonian and the single-particle density operator in the interaction picture after a canonical transformation, respectively. The external laser field is written as $\varepsilon(t) = \varepsilon_0 e^{-i\omega t}$, where $\omega$ is the center frequency of the incident laser and $\varepsilon_0$ is the peak field amplitude. The electron is assumed to occupy the Bloch state labeled by band index $\lambda$ and wave vector $k$.

Solving the above equation and neglecting dephasing, consistent with our TDDFT calculations, we compute the macroscopic polarization via $P(t) = tr\left[\rho(t)^{int} d^{int}\right]$, where $tr$ denotes the trace. The interband transition dipole element operator in the interaction picture reads $d_{vc}^{\mathrm{int}} = d_{vc}\, e^{i(\epsilon_{v,k} - \epsilon_{c,k})t}$, with $d_{vc} = d_{cv}^{*}$ being the optical transition dipole matrix element for excitation from the valence band to the conduction band. Here, $\epsilon_{v,k}$ and $\epsilon_{c,k}$ are the adiabatic eigenenergies of the valence- and conduction-band states at momentum $k$. Collecting terms, we obtain the time-dependent macroscopic polarization $P(t)$:

$$P(t) \cong \sum_k c_k \cdot \left[\frac{1}{n_k} - f_{c,k}(t)\right] \cdot \cos\,(\omega t) \tag{2}$$

Where $c_k = \frac{4}{V}\,\varepsilon_0\, d_{cv}^2(\Gamma)\, E_g^2(\Gamma)\, \frac{1}{E_g^3(k)} \propto \frac{\varepsilon_0}{E_g^3(k)}$, where $E_g(k)$ denotes the adiabatic band gap at wave vector $k$. The quantity $f_{c,k}(t)$ represents the time-dependent electron occupation in the conduction band, and $n_k$ is the number of $k$-points used in sampling, and $\Gamma$ denotes the Brillouin-zone center (i.e., $\mathbf{k} = 0$).

The time-dependent emission of the interband current $J_{\mathrm{er}}(t)$ is given by the second time derivative of the macroscopic polarization,

$$J_{er}(t) = \ddot{P}(t) \cong \sum_k c_k \cdot \left\{\left[-\ddot{f}_{c,k}(t) - \omega^2 \cdot \left(\frac{1}{n_k} - f_{c,k}(t)\right)\right] \cdot \cos(\omega t) + \left[(\omega + 1) \cdot \dot{f}_{c,k}(t) - 1\right] \cdot \sin(\omega t)\right\} \tag{3}$$

We now analyze the right-hand side of Eq. (3). Its temporal dependence can be understood as the combination of two ingredients: (i) terms governed by the time-dependent conduction-band occupation $f_{c,k}(t)$, and (ii) a field-synchronous (or field-asynchronous) term proportional to $\cos(\omega t)$ and $\sin(\omega t)$.

We first consider the $\cos(\omega t)$ and $\sin(\omega t)$ factors. The $\cos(\omega t)$ component follows the driving field, whereas $\sin(\omega t) = \cos(\omega t - \pi/2)$ reflects the phase shift introduced by time differentiation. This phase lag is analogous to the classical response of polarization currents: the field-induced polarization current lags behind the conduction current of free carriers that is in phase with the driving field. We then turn to the contribution associated with $f_{c,k}(t)$. We introduce a simplified physical picture of electron excitation under the applied optical field (see the Appendix C for details). In this picture, the excitation probability is primarily modulated by the field-induced variation of the band gap, and thus tracks the phase electric field: $\cos(\omega t)$. Taking one-time derivative yields an additional $\pi/2$ phase shift, i.e., a $\sin(\omega t) = \cos(\omega t - \pi/2)$ dependence, whereas taking two-time derivatives restores a $\cos(\omega t)$-like behavior. In addition, one must account for the cumulative nature of the occupation dynamics (see the Appendix F for details). Electrons initially away from the Brillouin-zone center must first oscillate in $k$-space toward the momentum where the instantaneous band gap reaches its minimum before they can be efficiently excited. Consequently, $f_{c,k}(t)$ exhibits a "pre-acceleration"-like behavior [34], and the summed occupation dynamics over all sampled momenta become delayed relative to the driving field. Therefore, the interband emission is temporally delayed with respect to the optical field, originating from both the time differentiation of the macroscopic polarization and the occupation delay induced by the generalized pre-acceleration process. In contrast, the intraband emission remains largely in phase with the driving field (see the Appendix D for details).

$$J_{ra}(t) = \frac{4}{n_k} e \cdot \sum_{l=0}^{\frac{\pi}{d}} \cdot \sum_{s=1}^{\infty} \sum_{n=1}^{\infty} \alpha_n \frac{2nd}{\hbar} \cdot (2s-1) \cdot \omega \cdot J_{2s-1}\left(n \frac{\omega_B}{\omega}\right) \cdot \cos(ndl) \cdot \cos[(2s-1)\omega t] \quad (4)$$

Where $e$ is the elementary charge, $\alpha_n$ denotes the tight-binding expansion coefficients, $d$ is the nearest-neighbor distance, and $\hbar$ is the reduced Planck constant. $J_{2s-1}$ is the Bessel function of the first kind, and $\omega_B = (e\varepsilon_0 d)/\hbar$ is the Bloch oscillation frequency. The index $l$ labels the uniformly sampled crystal momenta in the first Brillouin zone.

Accordingly, the temporal correspondence between the interband and intraband emissions and the driving field is illustrated in Fig. 1a. Evidently, such time-perspective asynchrony broadens the full width at half maximum of the attosecond bursts, thereby degrading the temporal resolution of attosecond emission. This behavior arises because a stronger incident laser field promotes more carriers to higher-energy states, which enhances the delay of interband emission and strengthens the attosecond chirp [43,44]. This trend is fully confirmed by our TDDFT simulations.

The capability of our TDDFT approach to simulate HHG in semiconductors has been well validated in prior studies [45,46]. We now present and discuss our TDDFT simulation results. We consider a laser pulse with a duration of ~25 fs, a central wavelength of 2883 nm, and zero carrier–envelope phase (see the Appendix A for details). The laser intensity is kept below $1.00\ \mathrm{TW/cm^2}$. By applying spectral filtering, we extract the strongest half-cycle attosecond burst [4,7], where the filtering window is set below the cutoff energy at the lowest intensity (4.3 eV). Using bulk silicon as prototype, we analyze how the properties of the strongest filtered half-cycle attosecond burst depend on the laser intensity, as shown in Fig. 1. As the laser intensity increases, the temporal center of the attosecond burst initially shifts rapidly away from the electric-field maximum and then exhibits a slower drift at higher intensities (Fig. 1d). This behavior indicates that at low intensities the field is insufficient to significantly populate the conduction band, so the emission is dominated by the intraband mechanism and remains nearly synchronized with the field maximum. As the field strength increases, the interband contribution becomes increasingly important, and the delayed emission associated with this mechanism leads to a growing temporal shift, in agreement with our theoretical expectations. Interestingly, as shown in Fig. 1c, the FWHM of the attosecond burst initially increases with intensity, as anticipated, but then decreases as the intensity is further raised. This counterintuitive behavior arises because at sufficiently high fields the interband emission becomes dominant; consequently, the attosecond burst timing is no longer governed by the competition between intra- and interband channels, but instead by the enhanced strength of the interband emission, which leads to a narrowing of the FWHM. The resulting non-monotonic dependence of the attosecond temporal resolution on laser intensity provides strong evidence that the microscopic HHG mechanisms critically shape the emitted attosecond bursts. This implies that, in the search for optimal solid-state attosecond sources, one must identify an optimal operating regime balancing the intra- and interband emission channels.

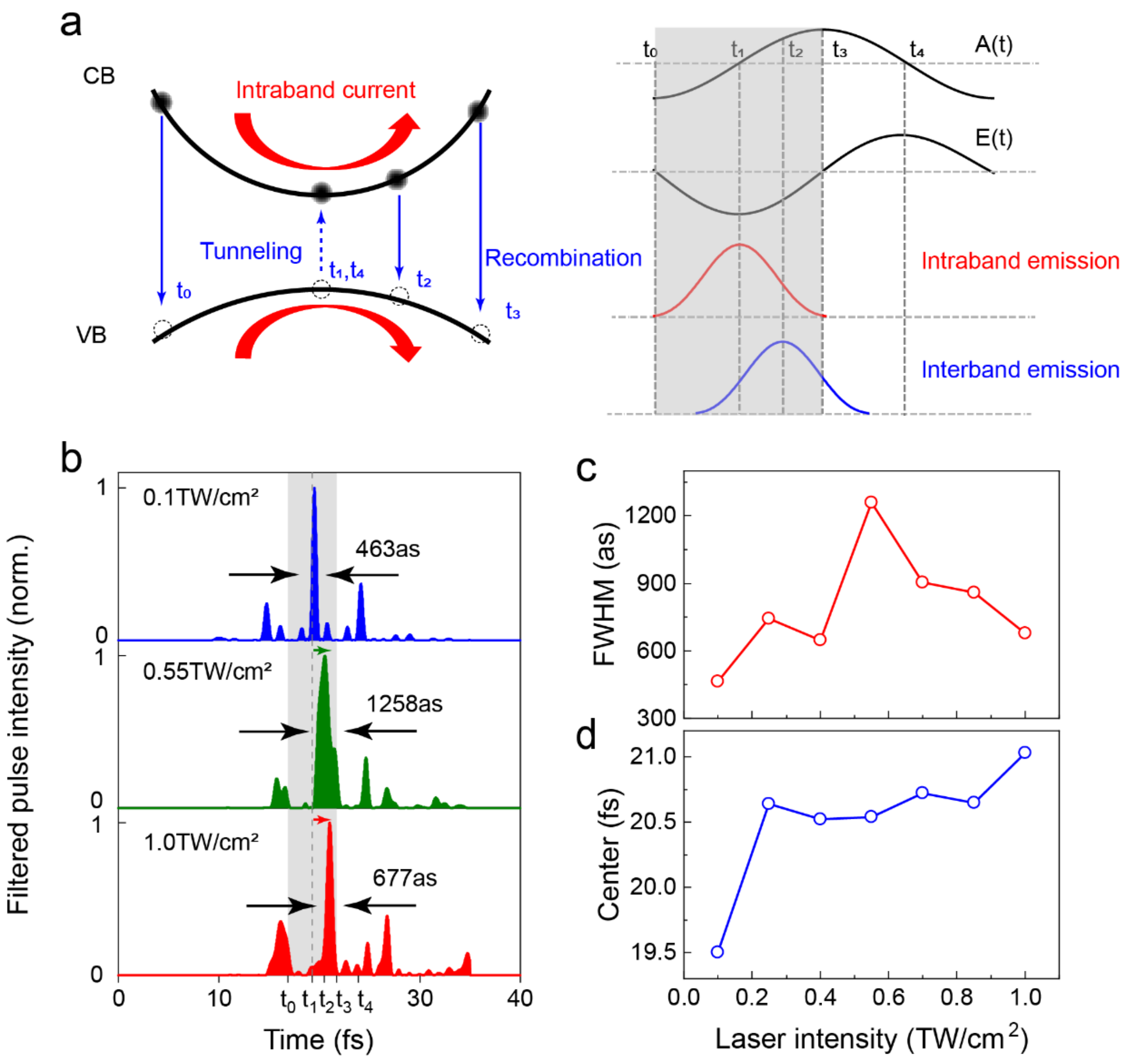

**FIG 1.** Dephased interband emission mechanism. (a) Schematic illustration of the temporal offset between the intraband and interband emission channels. (b) Intensity-dependent temporal profiles of attosecond pulse trains. The blue, green, and red shaded areas correspond to laser intensities of $0.10$, $0.55$, and $1.00$ $\mathrm{TW/cm^2}$, respectively. The gray shaded region marks the emission window of the strongest half cycle of the driving field, and the gray dashed line indicates the peak of the electric field within this window. The black arrow denotes FWHM of the strongest filtered attosecond burst, while the green and red arrows indicate the temporal offsets of the burst peak relative to the field maximum. (c) and (d) show the FWHM and the temporal center of the strongest attosecond burst as functions of the laser intensity, respectively.

We now examine how the time-dependent conduction-band occupation influences the temporal characteristics of the emitted attosecond bursts. From Eq. (3), the temporal structure of the macroscopic polarization, relative to the electric-field maximum, is governed by the time dependence of the conduction-band occupation. As discussed above, the temporal offset of the total attosecond emission with respect to the field arises exclusively from the interband channel; therefore, the timing of the attosecond bursts is ultimately influenced deeply by the dynamics of the conduction-band population.

Fig. 2 shows the time evolution of the main conduction-band occupations. At a low laser intensity of $0.10\,\mathrm{TW/cm^2}$, electrons are predominantly confined to the lowest conduction band (band 1), and the occupation dynamics are synchronized with the driving field, exhibiting a half-cycle periodicity. This behavior indicates that the emission is dominated by the intraband mechanism. When the intensity is increased to $0.55\,\mathrm{TW/cm^2}$, more electrons are promoted into the conduction bands and the interband contribution becomes significant. As electrons can be excited to higher-lying bands, the generalized pre-acceleration mechanism induces a delay in the occupation dynamics, as described above. This delay accumulates and becomes particularly pronounced in the highest conduction band (band 4), where a clear temporal lag of the population peak relative to the electric-field maximum is observed. At an even higher intensity of $1.00\ \mathrm{TW/cm^2}$, this effect is further enhanced, providing additional evidence that generalized pre-acceleration governs the delayed population dynamics (see the Appendix F for details). Consequently, the evolution of the conduction-band occupation provides a direct temporal fingerprint of the attosecond emission. This conclusion is further corroborated by an independent analysis based on the Heisenberg picture and an adiabatic-basis expansion [25] (see the Appendix E). Compared with conventional attosecond optical metrology techniques, such as RABBIT [47] and attosecond streaking [48], we propose that time-resolved probes of the Brillouin-zone population—such as time-resolved angle-resolved photoemission spectroscopy (trARPES) [49], Pauli blocking, and attosecond transient absorption [50–52]—can be used to directly access the temporal structure of attosecond emission. These approaches offer two key advantages. First, they enable momentum-resolved attribution in solids, allowing one to identify which bands or even which $k$ point dominates the emission process. Second, whereas previous studies relied on semiconductor Bloch equation (SBE) simulations combined with time–frequency analysis to infer the emission mechanisms indirectly [9,15,16,44,53], the present approach allows a direct discrimination between intra- and interband contributions based on the relative timing between the population dynamics and the driving field.

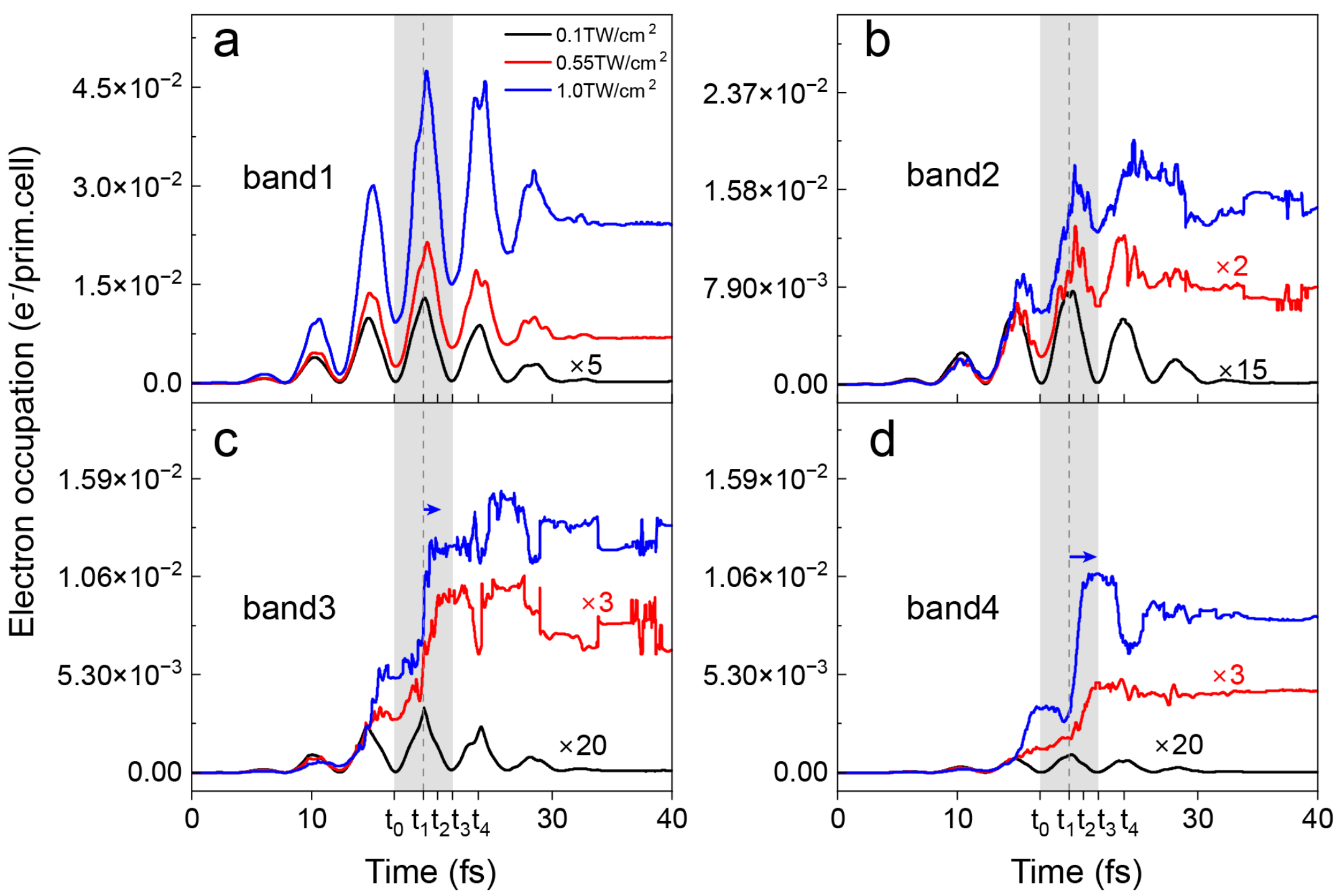

**FIG 2.** Laser-intensity-dependent conduction-band population dynamics. (a)–(d) Time evolution of the electron occupations at different laser intensities on band1 (the lowest conduction band), band2, band3 and band4, respectively. The black, red, blue solid line are $0.10, 0.55$ and $1.00\,\mathrm{TW/cm^2}$, respectively. The gray shaded region indicates the emission window of the strongest half cycle of the driving field, and the gray dashed line marks the field maximum within this window. Blue arrows denote the temporal offsets of the band3, band4 population peak relative to the electric-field maximum at different laser intensities.

To characterize the strong-field regime of our process, we evaluate key physical parameters of the emission dynamics [36]. Fig. 3 summarizes the strong-field ionization behavior at different laser intensities. In Fig. 3a, the blue solid line with data points shows the intensity dependence of the third harmonic, while the black dashed line represents a perturbative harmonic fit, indicating that the system operates in a nonperturbative regime. In Fig. 3b, we plot the number of excited electrons as a function of laser intensity. The Keldysh parameter $\gamma = \omega\sqrt{|\,m_{e-h}\,|\,E_g}/(e\varepsilon_0)$ is used to distinguish between the multiphoton ($\gamma \gg 1$) and tunneling ($\gamma \ll 1$) excitation

regimes [36,54], revealing that tunneling excitation dominates in our case. Here, $\omega$ is the central laser frequency, and $m_{e-h} = (m_e m_h)/(m_e + m_h)$ is the reduced effective mass of electrons and holes along the laser polarization direction. $E_g$ denotes the band gap and $\varepsilon_0$ the peak electric-field amplitude. In our first-principles calculations, $\omega = 0.0158$a.u. Since strong-field ionization predominantly occurs near the $\Gamma$ point [55], we use the electron and hole parameters at $\Gamma$ for simplicity, with $m_e = -0.466$a.u., $m_h = -0.273$a.u., and $E_g = 0.0936$a.u.

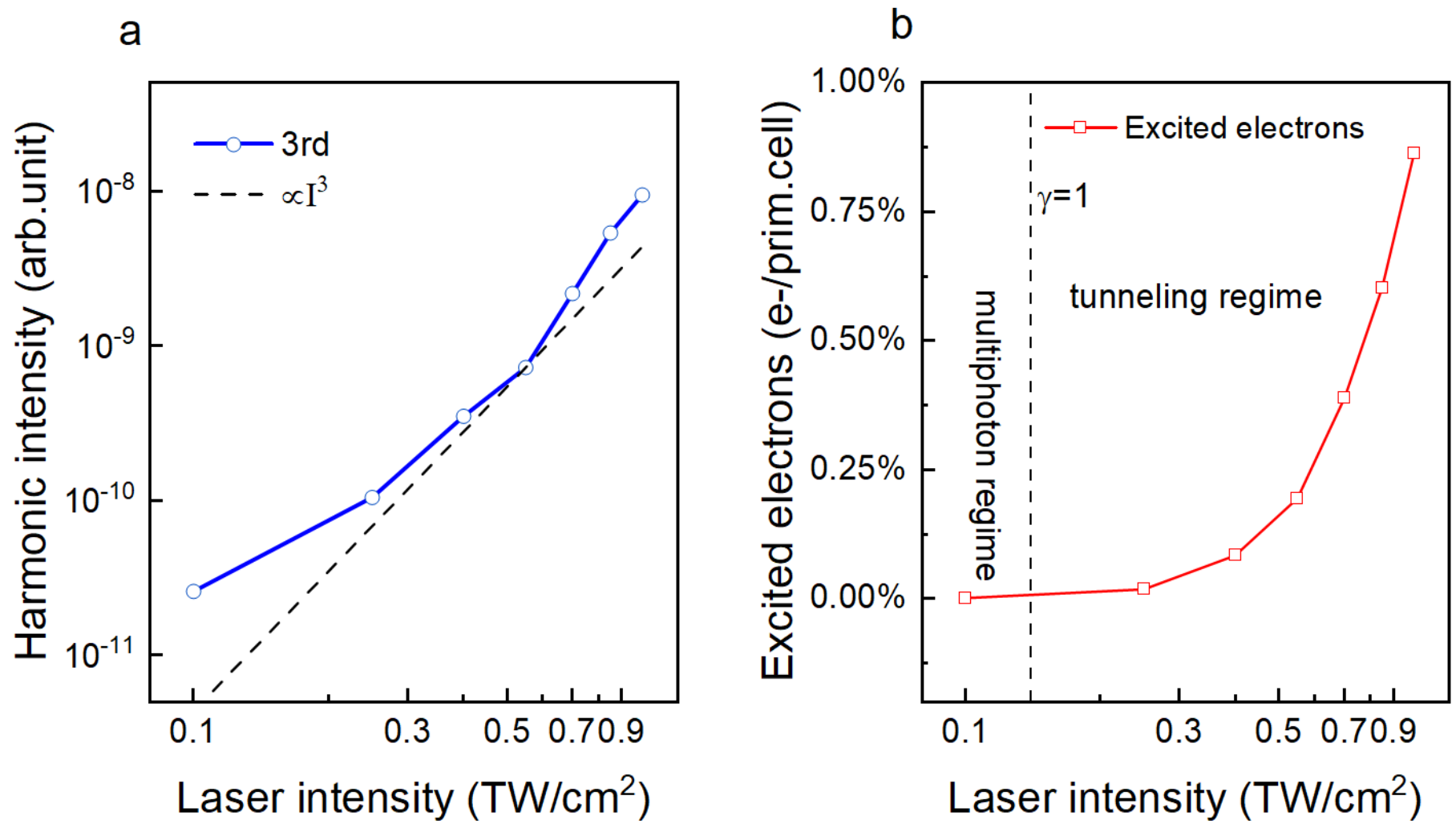

**FIG 3.** Strong-field ionization dynamics at different laser intensities. (a) Intensity dependence of the third-harmonic and its perturbative scaling. The blue solid line with data points shows the third-harmonic yield as a function of laser intensity, while the black dashed line represents the corresponding perturbative fit. (b) Intensity dependence of the number of excited electrons. The red solid line with data points shows the excited-electron population versus laser intensity. The black vertical dashed line indicates the boundary between the multiphoton and tunneling excitation regimes.

## IV. CONCLUSION

Together, these results have identified a fundamental time-domain limitation of solid-state attosecond emission that cannot be captured by conventional frequency-domain arguments based on harmonic cutoff extension. By establishing an occupation-resolved description of attosecond emission, interpreting through our proposing simplified excitation model and the generalized pre-acceleration picture, and validating it with real-time first-principles simulations, we demonstrate that interband emission is intrinsically delayed relative to field-synchronous intraband emission due to carrier-occupation dynamics. The resulting temporal mismatch between the two emission channels leads to a nonmonotonic dependence of attosecond pulse width on laser intensity, even well below material damage thresholds.

Our results reveal that, in solids, attosecond pulse compression is fundamentally constrained by carrier population dynamics rather than spectral bandwidth alone. This insight revises the prevailing intuition for optimizing solid-state attosecond sources and highlights the importance of explicitly accounting for time-domain synchronization between microscopic emission processes in strong-field light-matter interaction. Future work will focus on how to actively control the balance between interband and intraband contributions to optimize the properties of emitted attosecond pulses.

**ACKNOWLEDGEMENTS**

This work was supported by the National Natural Science Foundation of China (Grant No. 12174380).

**Supplemental Material of**

# Occupation-Driven emission asynchronous as a Fundamental Constraint on Solid-State Attosecond Pulses

## Appendix A. Computational details of the first-principles method

All static and real-time time-dependent density functional theory (rt-TDDFT) calculations were performed using the first-principles package PWMAT[35,55]. We employed optimized norm-conserving Vanderbilt pseudopotentials [40] together with the adiabatic Perdew–Burke–Ernzerhof (PBE) exchange–correlation functional. The time-evolving wave functions were expanded in a plane-wave basis set with a kinetic-energy cutoff of 50 Ry, and Brillouin-zone sampling was carried out using a $28 \times 28 \times 28$ Monkhorst–Pack $k$-point grid for the irreducible unit cell. Silicon $(2s^2 2p^2)$ electrons were treated as valence electrons.

It is well known that conventional DFT underestimates the band gap due to the absence of self-interaction corrections [38] and excitonic effects [56]. However, neglecting these effects does not qualitatively affect our conclusions. In addition, spin effects, dephasing due to lattice and impurity scattering [2], and finite carrier lifetimes were not included. Other nonlinear optical effects, such as self-focusing and propagation, were also neglected.

A primitive unit cell of silicon with a lattice constant of $a = 5.47$ Å was used. The driving laser field had a sine-squared envelope with a FWHM of 25 fs, a central wavelength of 2883 nm, and zero carrier–envelope phase. The peak intensity was varied from 0.1 to 1.0 $\mathrm{TW/cm^2}$ (in vacuum, neglecting surface reflections). The laser polarization was aligned along the [100] crystallographic direction.

## Appendix B. Analytical derivation of the delayed interband emission

An electron occupying the Bloch state labeled by band index $\lambda$ and crystal momentum $k$ satisfies the orthonormality condition:

$$\langle \lambda' k' | \lambda k \rangle = \delta_{\lambda\lambda'} \cdot \delta_{kk'} \tag{1}$$

$$\sum_{\lambda k} |\lambda k\rangle \langle \lambda k| = 1 \tag{2}$$

The time-dependent Schrödinger equation for an electron in the presence of the applied electric field $\boldsymbol{E(t)}$ reads:

$$i\hbar \frac{\partial \Psi(r,t)}{\partial t} = H \cdot \Psi(r,t) = (H_0 + H_I) \cdot \Psi(r,t) \tag{3}$$

$$H_0 = -\frac{\hbar^2 \nabla^2}{2m} + V(r) \tag{4}$$

$$H_I = -er \cdot \varepsilon(t) \tag{5}$$

Here, $V(\mathrm{r})$ denotes the periodic crystal potential. The field-free Hamiltonian $H_0$ satisfies the eigenvalue equation: $H_0 \mid \lambda k\rangle = E_\lambda(k) \mid \lambda k\rangle = \hbar\epsilon_{\lambda k} \mid \lambda k\rangle$, where $\boldsymbol{E_\lambda(k)} = \boldsymbol{\hbar\epsilon_{\lambda k}}$ is the corresponding eigenenergy. The interaction Hamiltonian $H_I$ is written in the length gauge, with $\varepsilon(t)$ being the applied electric field, $\varepsilon(t) = \varepsilon_0 e^{-i\omega t}$.

Substituting Eq. (2) into Eqs. (4) and (5) yields:

$$H_0 = \hbar \sum_{\lambda k} \epsilon_{\lambda k} |\lambda k\rangle \langle \lambda k| \tag{6}$$

$$H_I = -er \cdot \varepsilon(t) \sum_{k,k',\lambda,\lambda'} r_{\lambda,\lambda'}(k,k') \, |\lambda' k'\rangle\langle \lambda k| \tag{7}$$

Here, $r_{\lambda,\lambda'}(k,k') = \langle \lambda' k' | r | \lambda k\rangle$. It can be evaluated via

$$r_{\lambda,\lambda'}(k,k') = \frac{1}{E_\lambda(k) - E_{\lambda'}(k')} \langle \lambda' k' | [r, H_0] | \lambda k\rangle = \frac{i}{m(\epsilon_{\lambda k} - \epsilon_{\lambda' k'})} \langle \lambda' k' | p | \lambda k\rangle \tag{8}$$

Therefore, by invoking the $k \cdot p$ perturbation framework, the dipole matrix element $d$ satisfies

$$er_{\lambda,\lambda'}(k,k') = d_{\lambda,\lambda'}(k,k') = \delta_{kk'} \cdot d_{\lambda\lambda'}(0) \frac{\epsilon_{\lambda',0} - \epsilon_{\lambda,0}}{\epsilon_{\lambda',k} - \epsilon_{\lambda,k'}} \tag{9}$$

Substituting Eq. (9) into Eq. (7) yields:

$$H_I = -\varepsilon(t) \sum_{k,\{\lambda \neq \lambda'\} = c,v} d_{\lambda,\lambda'} \, |\lambda' k\rangle\langle \lambda k| \equiv \sum_k H_{I,k} \tag{10}$$

$$H_{I,k} = -\varepsilon(t) \cdot (d_{cv} |ck\rangle\langle vk| + {d_{cv}}^* |vk\rangle\langle ck|) \tag{11}$$

Here, $d_{cv}^* = d_{vc}$. For convenience in the subsequent calculations, we perform a canonical transformation on $H_{I,k}$:

$$H_{I,k}{}^{int}(t) = \exp\left(\frac{i}{\hbar}H_0\mathrm{t}\right)\cdot H_{I,k}\cdot\exp\left(-\frac{i}{\hbar}H_0\mathrm{t}\right) = -\varepsilon(t)\left[e^{i(\epsilon_{c,k}-\epsilon_{v,k})t}d_{cv}|ck\rangle\langle vk| + h.c.\right] \quad (12)$$

Here, $h.c.$ denotes the Hermitian conjugate of the preceding term.

For a single-particle density matrix $\rho(k)$:

$$\rho(k) = \sum_{\lambda'\lambda}\rho_{\lambda'\lambda}(k,t)\,|\lambda'k'\rangle\langle\lambda k| \quad (13)$$

In the interaction picture, the Liouville equation yields:

$$\frac{d}{dt}\,\rho_k^{int}(t) = -\frac{i}{\hbar}\left[H_{I,k}{}^{int},\rho_k^{int}(t)\right] \quad (14)$$

Here, $\rho_k^{int} = \exp\left(\frac{i}{\hbar}H_0\mathrm{t}\right)\cdot\rho(k)\cdot\exp\left(-\frac{i}{\hbar}H_0\mathrm{t}\right)$. Combining Eq. (11) with Eq. (14), we obtain:

$$\frac{d}{dt}\,\rho_k^{int}(t) = -\frac{i}{\hbar}\varepsilon(t)\sum_{\lambda,\lambda'}\rho_{\lambda'\lambda}{}^{int}(k,t)\times[e^{i(\epsilon_{c,k}-\epsilon_{v,k})t}d_{cv}(|ck\rangle\langle vk|\lambda'k\rangle\langle\lambda k| - |\lambda'k\rangle\langle\lambda k|ck\rangle\langle vk|)$$
$$+e^{-i(\epsilon_{c,k}-\epsilon_{v,k})t}d_{cv}{}^{*}(|vk\rangle\langle ck|\lambda'k\rangle\langle\lambda k| - |\lambda'k\rangle\langle\lambda k|vk\rangle\langle ck|)] \quad (15)$$

We evaluate the matrix element $\rho_{cv}^{\mathrm{int}}(t)$:

$$\rho_{cv}^{int}(k,t) = \left\langle ck\middle|\rho_k^{int}(t)\middle|vk\right\rangle \quad (16)$$

Combining with Eq. (15), we obtain:

$$\rho_{cv}^{int}(k,t) = \frac{i}{\hbar}d_{cv}\cdot\varepsilon(t)\cdot e^{i(\epsilon_{c,k}-\epsilon_{v,k})t}\cdot[\rho_{vv}(k,t) - \rho_{cc}(k,t)] \quad (17)$$

Here we have used the relation $\rho_{\lambda\lambda}^{\mathrm{int}} = \rho_{\lambda\lambda}$.

Similarly, for a multiband system the same conclusion remains valid. For convenience, we define:

$$\rho_{\lambda\lambda(k,t)} \equiv f_{\lambda,k}(t) \quad (18)$$

Substituting Eq. (17) and carrying out the time integration, we obtain:

$$\rho_{cv}^{int}(k,t) = \frac{d_{cv}}{\hbar}\cdot\varepsilon_0\cdot\frac{e^{i(\epsilon_{c,k}-\epsilon_{v,k}-\omega)t}}{\epsilon_{c,k}-\epsilon_{v,k}-\omega-i\gamma}\cdot\left[f_{v,k}(t) - f_{c,k}(t)\right] \quad (17)$$

Introducing a phenomenological damping parameter $\gamma$, though we now consider a strong field situation, whenever it works well in standard practice [42]. It is commonly motivated by writing the driving field as $\varepsilon(t) = \lim_{\gamma\to 0}\varepsilon_0\cdot\mathrm{e}^{-\mathrm{i}\omega\mathrm{t}}\cdot\mathrm{e}^{\gamma\mathrm{t}}$, which ensures that the field vanishes in the remote past ($t\to-\infty$) in the context of response theory. Physically, this corresponds to an adiabatic switch-on of the interaction, and it implicitly enforces causality and dynamical stability.

However, in our first-principles simulations we neglect dephasing, i.e., we set $\gamma = 0$, so that:

$$\rho_{cv}^{int}(k,t) = \frac{d_{cv}}{\hbar} \cdot \varepsilon_0 \cdot \frac{e^{i(\epsilon_{c,k}-\epsilon_{v,k}-\omega)t}}{\epsilon_{c,k}-\epsilon_{v,k}-\omega} \cdot \left[f_{v,k}(t) - f_{c,k}(t)\right] \tag{18}$$

The optical polarization is given by:

$$P(t) = tr[\rho(t)d] = tr\left[\rho(t)^{int} d^{int}\right] \tag{19}$$

Here, $tr$ denotes the trace. Collecting terms, we obtain:

$$P(t) = \frac{1}{V}\sum_k \left[\rho_{cv}^{int}(k,t) d_{vc}^{int}(k,t) + \rho_{vc}^{int}(k,t) d_{cv}^{int}(k,t)\right] = \frac{1}{V}\sum_k \frac{\varepsilon_0 |d_{cv}|^2}{\hbar} \frac{f_{v,k}(t) - f_{c,k}(t)}{\epsilon_{c,k}-\epsilon_{v,k}-\omega} e^{-i\omega t} + c.c. ($$

Here, $d_{vc}^{\mathrm{int}} = d_{vc}\, e^{i(\epsilon_{v,k}-\epsilon_{c,k})t}$, $c.c.$ denotes the complex conjugate of the preceding term, and $V$ is the crystal volume. Further collecting terms, we obtain:

$$P(t) = \frac{1}{V}\sum_k \varepsilon_0 |d_{cv}|^2 \cdot [f_{v,k}(t) - f_{c,k}(t)] \cdot [\frac{1}{\hbar(\epsilon_{c,k}-\epsilon_{v,k}-\omega)} - \frac{1}{\hbar(\epsilon_{v,k}-\epsilon_{c,k}-\omega)}] \cdot e^{-i\omega t} \tag{21}$$

The valence-band occupation can be expressed in terms of the conduction-band occupation as

$$f_{v,k}(t) = \frac{2}{n_k} - f_{c,k}(t) \tag{22}$$

Here, $n_k$ is the number of $k$-points used in the sampling. For convenience, we define $\hbar(\epsilon_{c,k} - \epsilon_{v,k}) \equiv E_g(k)$. Combining Eq. (22) with Eq. (21), we obtain

$$P(t) = \frac{4}{V}\sum_k \varepsilon_0 |d_{cv}|^2 \cdot [\frac{1}{n_k} - f_{c,k}(t)] \cdot \frac{E_g(k)}{E_g(k)^2 - \hbar^2\omega^2} \cdot e^{-i\omega t} \tag{23}$$

From Eq. (9), the dipole matrix element satisfies:

$$d_{cv}(k,k') = d_{cv}(k=0)\frac{E_g(k=0)}{E_g(k)} \tag{24}$$

Substituting Eq. (24) into Eq. (23) yields:

$$P(t) = \frac{4}{V}\sum_k \varepsilon_0 {d_{cv}}^2(k=0)\frac{{E_g}^2(k=0)}{{E_g}^2(k)} \cdot [\frac{1}{n_k} - f_{c,k}(t)] \cdot \frac{E_g(k)}{E_g(k)^2 - \hbar^2\omega^2} \cdot e^{-i\omega t} \tag{25}$$

Further rearranging terms, we obtain

$$P(t) = \frac{4}{V}\sum_{k} \varepsilon_0 {d_{cv}}^2(k=0) {E_g}^2(k=0) \cdot [\frac{1}{n_k} - f_{c,k}(t)] \cdot \frac{1}{(E_g(k)^2 - \hbar^2\omega^2)\cdot E_g(k)} \cdot e^{-i\omega t} \quad (26)$$

In high-harmonic generation, the driving wavelength is typically in the mid-infrared or far-infrared regime, such that $\hbar\omega \ll E_g(k)$. Under this condition, Eq. (26) can be simplified to

$$P(t) \cong \frac{4}{V}\sum_{k} \varepsilon_0 {d_{cv}}^2(k=0) {E_g}^2(k=0) \cdot \frac{1}{{E_g}^3(k)} \cdot [\frac{1}{n_k} - f_{c,k}(t)] \cdot e^{-i\omega t} \quad (27)$$

We define $c_k \equiv \frac{4}{V} \cdot \varepsilon_0 {d_{cv}}^2(k=0) {E_g}^2(k=0) \cdot \frac{1}{{E_g}^3(k)} \propto \frac{\varepsilon_0}{{E_g}^3(k)}$, so that Eq. (27) reduces to

$$P(t) \cong \sum_k c_k \cdot \left[\frac{1}{n_k} - f_{c,k}(t)\right] \cdot e^{-i\omega t} \quad (28)$$

Taking the real part yields the macroscopic polarization

$$P(t) \cong \sum_k c_k \cdot \left[\frac{1}{n_k} - f_{c,k}(t)\right] \cdot \cos{(\omega t)} \quad (29)$$

The interband current in the acceleration form is then obtained as

$$\dot{P}(t) \cong \sum_k c_k \cdot \left\{ -\dot{f}_{c,k}(t) \cdot \cos(\omega t) + \left[\frac{1}{n_k} - f_{c,k}(t)\right] \cdot -\omega \cdot \sin(\omega t) \right\} \quad (30)$$

$$J_{er}(t) = \ddot{P}(t) \cong \sum_k c_k \cdot \left\{ \left[ -\ddot{f}_{c,k}(t) - \omega^2 \cdot \left( \frac{1}{n_k} - f_{c,k}(t) \right) \right] \cdot \cos(\omega t) + \left[ (\omega + 1) \cdot \dot{f}_{c,k}(t) - 1 \right] \cdot \sin(\omega t) \right.$$

To obtain the population dynamics, we first consider only the excitation of conduction-band electrons by the driving field. For clarity, the time evolution of the electron occupation in reciprocal space is presented in **Appendix C**.

**Appendix C. Simplified physical picture of electron tunneling under an external optical field**

For simplicity, we consider the adiabatic tunneling regime [36]. We assume that the conduction band is initially empty. After excitation, the conduction- and valence-band occupations are denoted by $n_c$ and $n_v$, respectively. During the tunneling process, electrons exchange energy through electron–electron interactions in order to satisfy the dispersion relation, while conserving the total energy and momentum. The tunneling probability for an electron at crystal momentum $K$ and time $t$ can then be written as

$$prob(K,t) = \frac{n_c(K')}{n_c(K') + n_v(K')} \tag{32}$$

Where $K'$ denotes the crystal momentum of the electron after tunneling.

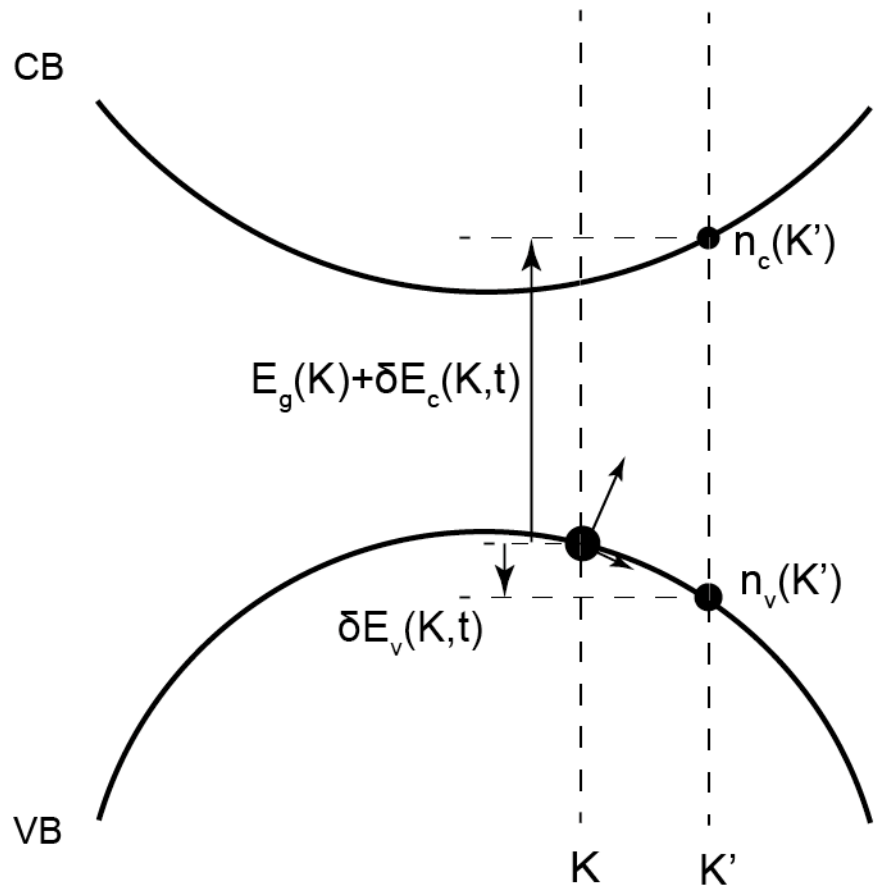

Figure S1. Schematic illustration of the simplified tunneling picture. Under the excitation of the external optical field, an electron undergoes adiabatic tunneling from the valence band (VB) at crystal momentum $K$ to the conduction band (CB) and the valence band at momentum $K'$. After excitation, the corresponding populations are $n_c(K')$and $n_v(K')$, respectively. The associated band-gap modifications are given by an increase $E_g(K) + \delta E_c(K,t)$ in the conduction band and a decrease $\delta E_v(K,t)$ in the valence band.

By energy conservation together with crystal-momentum conservation (energy conservation for the many-electron system and crystal-momentum conservation for a single electron), we obtain

$$-n_v(K') \cdot \delta E_v(K) = n_c(K',t) \cdot \left(E_g(K) + \delta E_c(K)\right) \tag{33}$$

Substituting Eq. (30) into Eq. (31) yields

$$prob(K',t) = -\frac{\delta E_v(K,t)}{E_g(K) + \delta E_c(K,t) - \delta E_v(K,t)} \cong -\frac{\delta E_v(K)}{E_g(K)} \quad (34)$$

Within the tight-binding approximation [39], the band dispersion can be written as

$$E_i(k) = \sum_R \varepsilon_i(R) \cdot e^{ikR} \quad (35)$$

And

$$\varepsilon_i(R) = \langle i,0|H_0|i,R\rangle = \frac{V}{(2\pi)^3}\int_{BZ} E_i(k) \cdot e^{-ikR} d^3k \quad (36)$$

$\varepsilon_i(R)$ is the hopping integral, i.e., the real-space matrix element of the field-free Hamiltonian $H_0$ between Wannier functions in band $i$ separated by the lattice vector $R$. Here, $V$ is the real-space primitive-cell volume, and $\varepsilon_i(0)$ corresponds to the band offset.

For convenience, we take $i = c, v$ to denote the conduction and valence bands, respectively. Substituting Eq. (35) into Eq. (34) and invoking the Bloch acceleration theorem $K'(t) = K + A(t)$, we obtain n

$$prob(K',t) = -\frac{\delta E_v(K)}{\sum_R [\varepsilon_c(R) - \varepsilon_v(R)] \cdot e^{ikR}} \quad (37)$$

For further analysis, we integrate on both sides:

$$\int_t^{t+dt} prob(K',t)\, dt = -\frac{\sum_R \varepsilon_v(R) \cdot iR \cdot e^{ikR} \cdot A(t)}{\sum_R [\varepsilon_c(R) - \varepsilon_v(R)] \cdot e^{ikR}} \quad (38)$$

We note that during tunneling, part of the electronic energy decreases while another part increases; here this energy redistribution is attributed to electron–electron interactions that enable the dispersion relation to be satisfied. Additionally, the above expressions are not restricted to a two-band model and can be generalized to multiband systems. The resulting tunneling probability depends explicitly on the band gap, band dispersion, and the driving field, consistent with the Landau–Zener tunneling picture [35].

We now revisit the right-hand side of Eq. (31), whose time dependence can be separated into two components: (i) a contribution controlled by the time-dependent conduction-band occupation $f_{c,k}(t)$, and (ii) a field-synchronous/asynchronous contribution involving $\cos(\omega t)$ and $\sin(\omega t)$. The $\cos(\omega t)$ term follows the driving field, whereas $\sin(\omega t) = \cos(\omega t - \pi/2)$ arises from the phase shift introduced by time differentiation.

For the $f_{c,k}(t)$-dependent part, we assume that the excitation probability is primarily modulated by the field-driven variation of the band gap and therefore follows a $\cos(\omega t)$-like temporal dependence. Taking one time derivative introduces an additional $\pi/2$ phase shift, yielding a $\sin(\omega t)$ dependence, whereas taking two-time derivatives restores a $\cos(\omega t)$-like response. In addition, one must account for the cumulative nature of the population dynamics. In the above tunneling picture, electrons initially away from the Brillouin-zone center must first oscillate in $k$-space toward the momentum where the instantaneous band gap reaches its minimum before they can be efficiently excited. Together with the analysis in **Appendix F**, this leads to a pre-acceleration-like behavior [34], such that the summed occupation dynamics over all $k$ points lag behind the driving field and introduce a temporal delay.

Consequently, the interband emission is delayed with respect to the optical field, originating from both the time differentiation of the macroscopic polarization and the occupation delay induced by the pre-acceleration-like dynamics.

**Appendix D. Analytical derivation of the absence of delay in intraband emission**

For simplicity, we first consider only the intraband current of conduction-band electrons. Owing to the inversion symmetry of the electron distribution after tunneling,

$$f_{c,k}(t) = f_{c,-k}(t) \tag{39}$$

The field-driven evolution of the crystal momentum is governed by the Bloch acceleration theorem,

$$\dot{k(t)} = -\frac{e}{\hbar} \cdot \varepsilon(t) \tag{40}$$

where the applied electric field is taken as $\varepsilon(t) = \varepsilon_0 \cdot \cos(\omega t)$. Consequently, the time-dependent conduction-band occupation can be written as

$$f_{c,k}(t) = f_{c,k-\Delta k}(t) \tag{41}$$

We consider two time-reversal-related k points in the Brillouin zone (assuming a spin-independent Hamiltonian and Kramers degeneracy), $K_+(k_x, k_y)$ and $K_-(-k_x, -k_y)$. The initial excited-electron distribution is taken as

$$Occ(k,0) = \frac{4}{n_k} \cdot \eta \cdot e \left[ \frac{\delta(k - K_+)}{2} + \frac{\delta(k - K_-)}{2} \right] \tag{42}$$

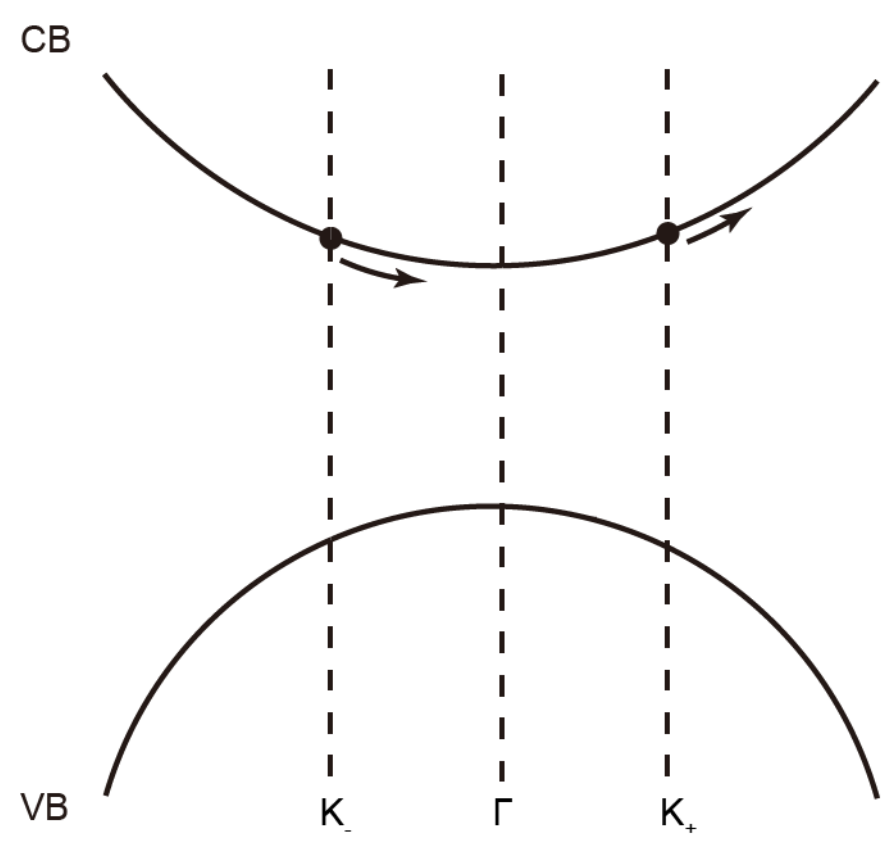

Figure S2. Schematic illustration of carrier-pair acceleration in the intraband process. Two inversion-symmetric crystal momenta $K_+(k_x, k_y)$and $K_-(-k_x, -k_y)$, initially located near the Brillouin-zone center, are accelerated by the driving laser field, as illustrated.

Using Eq. (41), this distribution evolves as

$$Occ(k,t) = \frac{4}{n_k} \cdot \eta \cdot e \left[ \frac{\delta\left(k - K_+ + \frac{e\varepsilon_0}{\hbar\omega} \cdot \sin(\omega t)\right)}{2} + \frac{\delta\left(k - K_- + \frac{e\varepsilon_0}{\hbar\omega} \cdot \sin(\omega t)\right)}{2} \right] \tag{43}$$

Here, $n_k$ is the number of $k$-points used in the sampling, $e$ is the elementary charge, and $\eta$ is the fraction of electrons instantaneously promoted to the conduction band.

We expand the conduction-band dispersion within a tight-binding form (keeping only the real part for clarity),

$$E_c(k) = \sum_{n=1}^{\infty} \alpha_n \cdot \cos(nkd) + E_g(k=0) \tag{44}$$

where $\alpha_n$ are tight-binding coefficients, $d$ is the nearest-neighbor distance, and $E_g(k=0)$ denotes the band gap at the Brillouin-zone center. The intraband current in the velocity form is given by

$$J_{ra,c}(t) = \sum_{k=0}^{\frac{\pi}{d}} Occ(k,t) \cdot \nabla_k E_c(k) \tag{45}$$

Substituting Eqs. (43) and (44) into Eq. (45), we obtain

$$J_{ra,c}(t) = \frac{4}{n_k} \cdot \eta \cdot e \cdot \sum_{l=0}^{\frac{\pi}{d}} \cdot \sum_{s=1}^{\infty} \sum_{n=1}^{\infty} \alpha_n \cdot \frac{2nd}{\hbar} \cdot J_{2s-1}\left(n\frac{\omega_B}{\omega}\right) \cdot \cos(ndl) \cdot \sin[(2s-1)\omega t] \tag{46}$$

where $J_{2s-1}$ is the Bessel function of the first kind and $\omega_B = (e\varepsilon_0 d)/\hbar$ is the Bloch frequency. Here, $l$ labels the uniformly spaced sampling points in the first Brillouin zone. Taking a time derivative yields the acceleration form,

$$J_{ra,c}(t) = \frac{4}{n_k} \cdot \eta \cdot e \cdot \sum_{l=0}^{\frac{\pi}{d}} \cdot \sum_{s=1}^{\infty} \sum_{n=1}^{\infty} \alpha_n \frac{2nd}{\hbar} \cdot (2s-1) \cdot \omega \cdot J_{2s-1}\left(n\frac{\omega_B}{\omega}\right) \cdot \cos(ndl) \cdot \cos[(2s-1)\omega t] \tag{47}$$

Since the intraband current oscillates in phase with $\cos[(2s-1)\omega t]$, its peak positions are synchronized with the driving field. A similar derivation shows that the intraband current associated with valence-band holes is also in phase with the field. Therefore, intraband emission is temporally synchronized with the optical field.

Notably, the above derivation accounts for the contribution of carriers over the entire Brillouin zone, regardless of whether they reach the minimum band-gap point. This is fully consistent with our simplified excitation and generalized pre-acceleration picture for interband emission, and it is not restricted to the scenario in which electrons must be accelerated to the minimum band gap before contributing to radiation [10,33,34].

## Appendix E. Theoretical derivation of attosecond pulse trains in the Heisenberg picture and the adiabatic-basis expansion

A couple of years ago, Nicolas Tancogne-Dejean et. al derived the approximate analytical expressions in ab initio calculation for HHG and APs within the single particle approximation using the language of second quantization [2]. Here, we have derived equivalent result based on the quantum mechanical framework in Heisenberg picture and extend it. For simplicity, we ignore the effect of nuclear motion [31,32], non-adiabatic Coulomb interaction of Hartree part and exchange-correlation functional [2,30,32], non-dipole approximation [29], spin polarization [28], electron and impurity scattering on Hamiltonian.

A physical quantity operator $\widehat{\boldsymbol{Q}}$ evolves over time according to the following equation;

$$\frac{\partial}{\partial t}\langle\widehat{\boldsymbol{Q}}\rangle = i\langle[\widehat{\boldsymbol{H}},\widehat{\boldsymbol{Q}}]\rangle + \langle\frac{\partial\widehat{\boldsymbol{Q}}}{\partial t}\rangle \tag{48}$$

The expression for the current in velocity can be written as follows,

$$\begin{aligned}\boldsymbol{j}(\boldsymbol{r},t) &= \frac{i}{2}\left(\Psi(\boldsymbol{r},t)\nabla\Psi(\boldsymbol{r},t)^* - \Psi(\boldsymbol{r},t)^*\nabla\Psi(\boldsymbol{r},t)\right) + \widehat{\boldsymbol{A}}(\boldsymbol{r},t)\cdot\Psi(\boldsymbol{r},t)\Psi(\boldsymbol{r},t)^* \\ &= \frac{1}{2}\left(\langle\Psi(\boldsymbol{r},t)|\widehat{\boldsymbol{p}} + \widehat{\boldsymbol{A}}(\boldsymbol{r},t)|\Psi(\boldsymbol{r},t)\rangle + c.c.\right)\end{aligned} \tag{49}$$

where $\Psi(\boldsymbol{r},t)$ is the wave function from solution of Kohn-Sham equation. $\widehat{\boldsymbol{A}}(\boldsymbol{r},t)$ is the operator of vector potential, $\widehat{\boldsymbol{p}}$ is the operator of momentum, $c.c.$ means the corresponding complex conjugate.

We take $\widehat{\boldsymbol{Q}} = \widehat{\boldsymbol{p}} + \widehat{\boldsymbol{A}}$, in velocity, consider the effectiveness of electron interaction and exchange-correlation term of ground state to Hamiltonian [2] and substituting it into equation（48）.

$$\begin{aligned}\frac{\partial}{\partial t}\langle\widehat{\boldsymbol{p}} + \widehat{\boldsymbol{A}}\rangle &= i\,\langle\frac{\left(\widehat{\boldsymbol{p}} + \widehat{\boldsymbol{A}}\right)^2}{2} + v_{eff}(\boldsymbol{r}),\widehat{\boldsymbol{p}} + \widehat{\boldsymbol{A}}\rangle + \langle\frac{\partial}{\partial t}(\widehat{\boldsymbol{p}} + \widehat{\boldsymbol{A}})\rangle \\ &= i\langle[v_{eff}(\boldsymbol{r}),\widehat{\boldsymbol{p}} + \widehat{\boldsymbol{A}}]\rangle + \langle\frac{\partial}{\partial t}(\widehat{\boldsymbol{p}})\rangle + \langle\frac{\partial}{\partial t}(\widehat{\boldsymbol{A}})\rangle \\ &= i\langle[v_{eff}(\boldsymbol{r}),\widehat{\boldsymbol{p}} + \widehat{\boldsymbol{A}}]\rangle + \langle\frac{\partial}{\partial t}(\widehat{\boldsymbol{A}})\rangle\end{aligned} \tag{50}$$

where $v_{eff}(\boldsymbol{r})$ includes electron-ion Coulomb potential $v_{ion}(\boldsymbol{r})$, electron interaction of Hartree part $v_{HF}(\boldsymbol{r})$ and exchange-correlation $v_{xc}(\boldsymbol{r})$ of ground state.

The first term of equation (50) can be simplified as follows,

$$i\langle[v_{eff}(\boldsymbol{r}),\widehat{\boldsymbol{p}} + \widehat{\boldsymbol{A}}]\rangle = i\langle[v_{eff}(\boldsymbol{r}),\widehat{\boldsymbol{p}}]\rangle = \langle\Psi(\boldsymbol{r},t)|v_{eff}\nabla - \nabla v_{eff}|\Psi(\boldsymbol{r},t)\rangle = -n(\boldsymbol{r},t)\nabla v_{eff}(\boldsymbol{r}) \tag{51}$$

where $n(\boldsymbol{r},t) = \Psi(\boldsymbol{r},t)^*\Psi(\boldsymbol{r},t)$ is the temporal electron density in space.

Through the relationship $\mathbf{E}(\mathbf{r},\mathrm{t}) \approx \mathbf{E}(\mathrm{t}) = -\partial\boldsymbol{A}(t)/\partial t$, we can obtain the electric field in dipole approximation and the second term can be written as follows.

$$\langle\frac{\partial}{\partial t}\left(\widehat{\boldsymbol{A}}\right)\rangle = -\langle\widehat{\boldsymbol{E}}(\boldsymbol{r},t)\rangle = -\left\langle\Psi(\boldsymbol{r},t)\middle|\widehat{\boldsymbol{E}}(\boldsymbol{r},t)\middle|\Psi(\boldsymbol{r},t)\right\rangle = -n(\boldsymbol{r},t)\cdot\boldsymbol{E}(t) \tag{52}$$

Substituting equations (51) and (52) into equation (50), we obtain

$$\langle\frac{\partial}{\partial t}\left(\widehat{\boldsymbol{p}}+\widehat{\boldsymbol{A}}\right)\rangle = -n(\boldsymbol{r},t)\cdot\left(\nabla v_{eff}(\boldsymbol{r})+\boldsymbol{E}(t)\right) \tag{53}$$

Thus, we derive the expression of APs and HHG,

$$\begin{aligned}
APT(\boldsymbol{r},t) &\propto \left|\int_{\Omega} d^3\boldsymbol{r}\,\left(\frac{\partial}{\partial t}\boldsymbol{J}(\boldsymbol{r},t)\right)\right|^2 \\
&\propto \left|\int_{\Omega} d^3\boldsymbol{r}\,\left(\frac{1}{2}\left(\left\langle\Psi(\boldsymbol{r},t)\middle|\widehat{\boldsymbol{p}}+\widehat{\boldsymbol{A}}(\boldsymbol{r},t)\middle|\Psi(\boldsymbol{r},t)\right\rangle + c.c.\right)\right)\right|^2 \\
&\propto \left|\int_{\Omega} d^3\boldsymbol{r}\,\left\{-n(\boldsymbol{r},t)\cdot\left(\nabla v_{eff}(\boldsymbol{r})+\boldsymbol{E}(t)\right)\right\}\right|^2 \\
&\propto \left|\int_{\Omega} d^3\boldsymbol{r}\left\{n(\boldsymbol{r},t)\nabla v_{eff}(\boldsymbol{r})+N_e(t)\right\}\right|^2
\end{aligned} \tag{54}$$

$$HHG(\omega) \propto \left|\mathcal{FT}\left\{\frac{\partial}{\partial t}\int_{\Omega} d^3\boldsymbol{r}\,n(\boldsymbol{r},t)\nabla v_{eff}(\boldsymbol{r})+N_e\boldsymbol{E}(t)\right\}\right|^2 \tag{55}$$

Where $N_e$ is the number of total electrons in system $\Omega$, $\mathcal{FT}$ means Fourier transform. We could obtain HHG from the corresponding Fourier transform, consider electron interaction of Hartree part and exchange-correlation contributions are from internal forces, so equation (4) and (5) reduces to

$$APT(t) \propto \left|\int_{\Omega} d^3\boldsymbol{r}\,n(\boldsymbol{r},t)\nabla v_{ion}(\boldsymbol{r})+N_e\boldsymbol{E}(t)\right|^2 \tag{9}$$

$$HHG(\omega) \propto \left|\mathcal{FT}\left\{\frac{\partial}{\partial t}\int_{\Omega} d^3\boldsymbol{r}\,n(\boldsymbol{r},t)\nabla v_{ion}(\boldsymbol{r})+N_e\boldsymbol{E}(t)\right\}\right|^2 \tag{10}$$

Noting equation (55) is equal to expression of literature [2], now we further write firstly $n(\boldsymbol{r},t)$ with adiabatic state basis expansion following PWMAT [25].

$$
\begin{aligned}
n(r,t) &= \sum_{i} n_i(r,t) \\
&= \sum_{i} \sum_{\boldsymbol{k}(t)\in BZ} OCC_{i,\boldsymbol{k}(t)} \Psi^*_{i,\boldsymbol{k}(t)}(\boldsymbol{r},t)\Psi_{i,\boldsymbol{k}(t)}(\boldsymbol{r},t) \\
&= \sum_{i} \sum_{\boldsymbol{k}(t)\in BZ} OCC_{i,\boldsymbol{k}(t)} \sum_{l} C^*_{i,l}(t) C_{i,l}(t) \varphi^*_{l,\boldsymbol{k}(t)}(t)\varphi_{l,\boldsymbol{k}(t)}(t) \\
&= \sum_{i} \sum_{\boldsymbol{k}(t)\in BZ} OCC_{i,\boldsymbol{k}(t)} \sum_{l,\boldsymbol{k}(t)\in BZ} \lbrace C^*_{i,l}(t) C_{i,l}(t) \left(\varphi^*_{l,\boldsymbol{k}(t)}(t)\varphi_{l,\boldsymbol{k}(t)}(t)\right) \\
&= \sum_{l,k} \widetilde{OCC}_{l,k(t)}(t) \cdot \varphi^*_{l,\boldsymbol{k}(t)}(t)\varphi_{l,\boldsymbol{k}(t)}(t)
\end{aligned}
\tag{56}
$$

Where $n_i(\boldsymbol{r},t)$ is the time-dependent electron spatial density which can be calculated through $n(r,t) = |\Psi_i(r,t)|^2$, $i$ is the index of the Kohn-Sham orbital, $\Psi_i(r,t)$ is the wavefunction, $OCC_{i,\boldsymbol{k}(t)}$ is the static electron occupation, $\widetilde{OCC}_{l,k(t)}(t)$ is time-dependent electron occupation, $\varphi_l$ is the adiabatic expansion basis, $l$ is the index of adiabatic basis.

At last, we substitute equation (56) into equations (54), (55) and obtain the final expression;

$$
\begin{aligned}
APs(t) &\propto \left| \int_{\Omega} d^3\boldsymbol{r} \sum_{i} n_i(\boldsymbol{r},t) \nabla\big(v_{ion}(\boldsymbol{r})\big) + N_e \boldsymbol{E}(t) \right|^2 \\
&\propto \left| \int_{\Omega} d^3\boldsymbol{r} \left( \sum_{l,k} \widetilde{OCC}_{l,k(t)}(t) \cdot \varphi^*_{l,\boldsymbol{k}(t)}(t)\varphi_{l,\boldsymbol{k}(t)}(t) \right) \cdot \nabla v_{ion}(\boldsymbol{r}) + N_e \boldsymbol{E}(t) \right|^2
\end{aligned}
\tag{57}
$$

$$
HHG(\omega) \propto \left| \mathcal{FT} \left\{ \int_{\Omega} d^3\boldsymbol{r} \left( \sum_{l,k} \widetilde{OCC}_{l,k(t)}(t) \cdot \varphi^*_{l,\boldsymbol{k}(t)}(t)\varphi_{l,\boldsymbol{k}(t)}(t) \right) \cdot \nabla v_{ion}(\boldsymbol{r}) + N_e \boldsymbol{E}(t) \right\} \right|^2 \tag{58}
$$

From the equations above, we can find the temporal occupation of carriers $\left(\widetilde{OCC}_{l,k(t)}(t)\right)$ have an obvious impact on the temporal characteristic of APs and HHG.

**Appendix F. Time evolution of the electron occupation under the generalized pre-acceleration process**

From Eq. (34), we see that the excitation probability under the vector potential $A(t)$ is primarily determined by the minimum band gap that an electron can reach during its laser-driven acceleration. Therefore, we focus on the time at which an electron attains the minimum band gap along its trajectory in $k$-space.

We consider half-cycle emission of the attosecond pulse train, i.e., $t \in [0, T/2]$, where $T$ is the laser period. For simplicity, invoking the Kramers degeneracy discussed in **Appendix D**, we analyze only electrons in the positive half of the first Brillouin zone, $k \in [0, \pi/a]$, where $a$ is the nearest-neighbor lattice spacing along the laser polarization direction. Without loss of generality, we assume that the vector potential satisfies $A(t) \in [0, A_{\max}]$ within a half cycle. We then classify electrons according to their initial crystal momentum and the maximum momentum displacement $A_{\max}$ that can be reached during the half cycle.

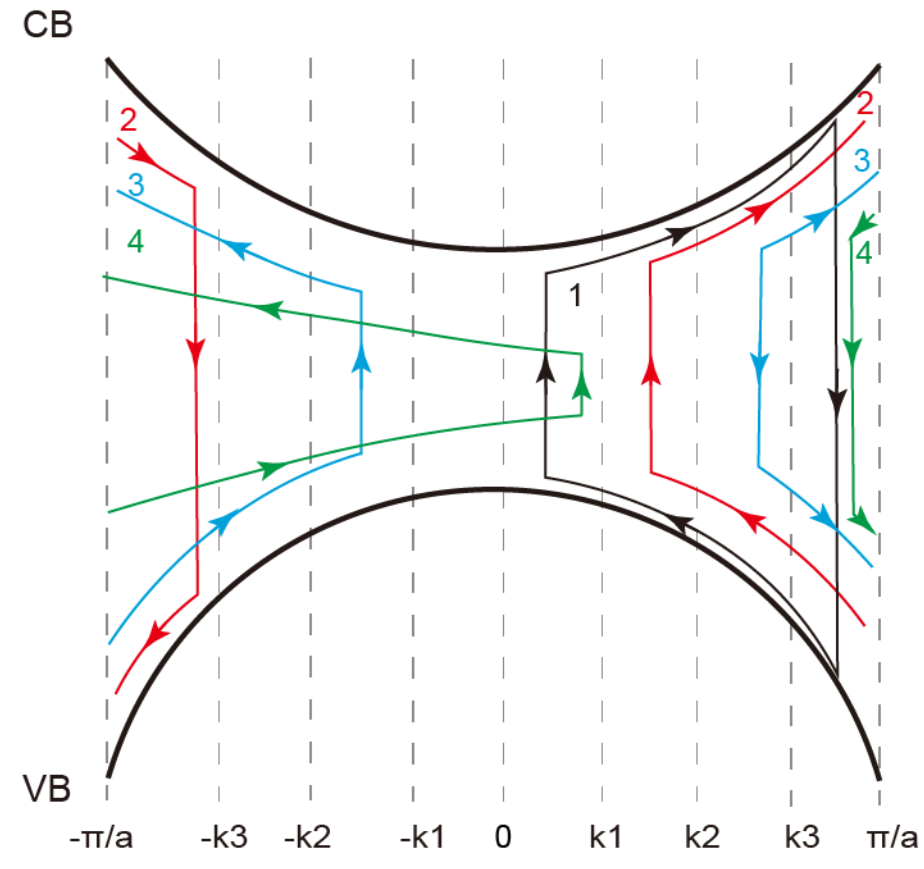

**Figure S3.** Schematic illustration of the generalized pre-acceleration picture for electron excitation. Electrons in the right half of the first Brillouin zone are classified according to their initial distance from the zone boundary. Region I ($0 \leq k < k_1$): electrons are farthest from the right Brillouin-zone boundary, and their maximum accelerated momentum remains inside the first Brillouin zone, as indicated by the black trajectory. Region II ($k_1 < k \leq k_2$): electrons are closer to the right zone boundary; although their maximum accelerated momentum exceeds the Brillouin-zone boundary, the minimum energy during the entire acceleration process still occurs at the initial momentum, as illustrated by the red trajectory. Region III ($k_2 < k \leq k_3$): electrons are even closer to the zone boundary; their maximum accelerated

momentum exceeds the first Brillouin zone, and the minimum energy during the trajectory shifts to the maximum momentum, as shown by the blue trajectory. Region IV ($k_3 < k \leq \pi/a$): electrons are closest to the right Brillouin-zone boundary and cross it immediately upon acceleration; similarly, the minimum energy along the trajectory occurs at the maximum momentum, as indicated by the green trajectory.

1. Region I ($0 \leq k \leq k_1$).

As indicated by the black trajectory, electrons reach the minimum band gap at $t = 0$, so that excitation to the conduction band occurs immediately and $f_{c,k}(t)$ exhibits a peak at $t = 0$. This requires $\mathrm{k} + A_{max} \leq \frac{\pi}{a}$, i.e., $0 \leq k \leq \frac{\pi}{a} - A_{\max}$.

2. Region II ($k_1 < k \leq k_2$).

As illustrated by the red trajectory, electrons can cross the first Brillouin-zone boundary, but the minimum band gap along the entire trajectory is still located at the initial momentum. Therefore, excitation again occurs at $t = 0$, leading to a peak in $f_{c,k}(t)$ at $t = 0$. This region is defined by $\frac{\pi}{a} < k + A_{max} \leq \frac{\pi}{a} + \frac{\pi}{a} - k_2$, which gives $\frac{\pi}{a} - A_{max} < k \leq \frac{\pi}{a} - \frac{A_{max}}{2}$.

3. Region III ($k_2 < k \leq k_3$).

As shown by the blue trajectory, electrons cross the first Brillouin zone and the minimum band gap along the trajectory shifts to the point of maximum crystal momentum. Consequently, excitation occurs at $t = T/4$, and $f_{c,k}(t)$ peaks at this time. The corresponding condition is $\frac{\pi}{a} < \mathrm{k} + A_{max} \leq \frac{\pi}{a} + \frac{\pi}{a}$, which yields $\frac{\pi}{a} - \frac{A_{max}}{2} < k \leq \frac{2\pi}{a} - A_{max}$.

4. Region IV ($k_3 < k \leq \pi/a$).

As indicated by the green trajectory, electrons cross the Brillouin-zone boundary immediately upon acceleration, and the minimum band gap is reached before the maximum momentum. So, $k > \frac{2\pi}{a} - A_{max}$Therefore, excitation occurs at $t < T/4$, leading to a delayed peak in $f_{c,k}(t)$.

The fourth region exists only when $\frac{\pi}{a} - A_{max} < 0$, i.e., $A_{max} > \frac{\pi}{a}$.

Taking all cases into account, the mean excitation time of the electron population, $t_{\mathrm{sum}}$, lies within the interval $t_{\mathrm{sum}} \in (0, T/4)$, implying a finite delay relative to the electric-field maximum at $t = 0$. Moreover, as the field strength increases (i.e., as $A_{\mathrm{max}}$becomes larger), the interval $|\pi/a - k_2|$ expands, so that Regions III and IV, together with the additional cases, occupy a larger fraction of the Brillouin zone. As a result, the excitation delay becomes more pronounced.

Finally, as shown in Eq. (38) of **Appendix B**, although the band gap is the dominant factor, the excitation probability also depends on the field strength and the detailed band dispersion. Here we focus on the primary role of the band gap, while the influence of these additional factors will be investigated in future work.